\documentclass[
    final            
  ]
  {aipproc}

\layoutstyle{6x9}

\newcommand{\hMpc}{{\ifmmode{h^{-1}{\rm Mpc}}\else{$h^{-1}$Mpc}\fi}}
\newcommand{\hkpc}{{\ifmmode{h^{-1}{\rm kpc}}\else{$h^{-1}$kpc}\fi}}
\newcommand{\hMsun}{{\ifmmode{h^{-1}{\rm {M_{\odot}}}}\else{$h^{-1}{\rm{M_{\odot}}}$}\fi}}
\newcommand{\Msun}{{\ifmmode{{\rm {M_{\odot}}}}\else{${\rm{M_{\odot}}}$}\fi}}

\begin{document}

\title{Dark and baryonic matter in the MareNostrum Universe}


\author{S. Gottl\"ober}{ address={Astrophysical Institute
    Potsdam, An der Sternwarte 16, 14482 Potsdam, Germany} }

\author{G. Yepes}{ address={Grupo de Astrof\'\i sica,
    Universidad Aut\'onoma de Madrid, Madrid E-28049, Spain} }

\author{A. Khalatyan}{ address={Astrophysical Institute
    Potsdam, An der Sternwarte 16, 14482 Potsdam, Germany} }

\author{R. Sevilla}{ address={Grupo de Astrof\'\i sica,
    Universidad Aut\'onoma de Madrid, Madrid E-28049, Spain} }

\author{V. Turchaninov}{
  address={Keldysh Institute for Applied Mathematics,
Miusskaja Ploscad 4,  125047 Moscow, Russia} }

\begin{abstract}
  We report some results from one of the largest hydrodynamical
  cosmological simulations of large scale structures that has been
  done up to date.  The {\em MareNostrum Universe} SPH simulation
  consists of 2 billion particles ($2\times 1024^3$) in a cubic box of
  500 $h^{-1}$ Mpc on a side.  This simulation has been done in the
  MareNostrum parallel supercomputer at the Barcelona SuperComputer
  Center. Due to the large simulated volume and good mass resolution,
  our simulated catalog of dark matter halos comprises more than half
  a million objects with masses larger than a typical Milky Way galaxy
  halo.  From this dataset we have studied several statistical
  properties such as the halo mass function, the distribution of
  shapes of dark and gas components within halos, the baryon fraction,
  cumulative void volume etc.  This simulation is particularly useful
  to study the large scale distribution of baryons in the universe as
  a function of temperature and density. In this paper we also show
  the time evolution of the gas fractions at large scales.
\end{abstract}

\maketitle


\section{Introduction}
During the last couple of decades the exciting observational
developments have enormously increased our knowledge about the history
of the universe.  A comparable progress has been made in our
theoretical understanding of the main processes that govern the
evolution of structure in the Universe.  A substantial part of this
progress is due to the increasing possibilities to simulate the
formation and evolution of structure on different scales using the new
generation of massive parallel supercomputers.   

The standard model of cosmological structure formation is based on the
idea of an early inflationary phase of the evolution of the Universe.
According to the simplest models of inflation during this phase
fluctuations with a scale free power spectrum have been created. On
large scales this power spectrum has been observed with high accuracy
by the WMAP satellite \cite{wmap} which measured the
fluctuations in the Cosmic Microwave Background radiation. The
cosmological models are  characterized by only five parameters: the
current rate of universal expansion, $H_0$, the mass density
parameter, $\Omega_{\rm mat}$, the value of the cosmological constant,
$\Omega_{\Lambda}$, the primordial baryon abundance, $\Omega_b$, and
the overall normalization of the power spectrum of initial density
fluctuations, typically characterized by $\sigma_8$, the present-day
rms mass fluctuations on spheres of radius 8$h^{-1}$ Mpc. 

Since 85 \% of the matter consists of dark matter particles the
gravitational evolution of the structures in the universe is dominated
by the dark matter. Many codes follow only the dark matter clumping
during cosmological evolution (e.g. \cite{millenium}). But most of the
information we get from the Universe comes from the baryons (either in
the form of X-ray emitting gas, or from the stars). If one is
interested in studying the distribution of baryons on large scales
then gas dynamics must be added to the gravitational evolution in a
cosmological simulation. This is not a trivial issue because
gasdynamical processes are very costly to simulate once the gas is
compressed to high densities when it falls into the potential wells of
the dark matter distribution.

Thanks to the recent advances in massively parallel computing, and to
the development of efficient MPI parallel codes that can use the total
computing power of thousand of processors linked together, numerical
simulations can treat more and more particles to describe the two main
components of the universe (collisionless dark matter and gas). This
allows to simulate larger and larger computational volumes with enough
resolution to identify typical galaxies like the Milky Way.

To compare the large scale distribution of dark matter and gas we have
been able to perform one of the largest cosmological gasdynamical
simulations ever done so far. We have followed the nonlinear evolution
of both the dark matter and the gas component within an 'adiabatic',
namely non-radiative and non-dissipative, cosmological Smooth Particle
Hydrodynamical (SPH) simulation. In what follows we will describe the
main features of the numerical simulation and will discuss some of the
results of the analyses we are doing in the large simulated dataset.

\section{Numerical simulation}

In our numerical simulation we have assumed the spatially flat
concordance cosmological model with the parameters $\Omega_m = 0.3$,
$\Omega_{bar} = 0.045$, $\Omega_{\Lambda} = 0.7$, the normalization
$\sigma_8 = 0.9$ and the slope $n=1$ of the power spectrum.  Within a
box of $500 \hMpc$ size the linear power spectrum at redshift $z=40$
has been represented by $1024^3$ DM particles of mass $m_{\rm DM} =
8.3 \times 10^{9} h^{-1} \Msun $ and $1024^3$ gas particles of mass
$m_{\rm gas} = 1.5 \times 10^{9} h^{-1} \Msun $. The nonlinear
evolution of structures has been followed by the GADGET II code of V.
Springel \cite{gadget}.  For the gravitational evolution we have used
the TREEPM algorithm on a homogeneous Eulerian grid to compute large
scale forces by the Particle-Mesh algorithm.  In this simulation we
employed $1024^3$ mesh points to compute the density field from
particle positions and FFT to derive gravitational forces.  Since the
baryonic component is also discretized by the gas particles all
hydrodynamical quantities have to be determined using interpolation
from the gas particles. Within GADGET the equations of gas dynamics
are solved by means of the Smoothed Particle Hydrodynamics method in
its entropy conservation scheme.  To follow structure formation until
redshift $z=0$ we have restricted ourselves to the gas-dynamics
without including dissipative or radiative processes or star
formation.  The spatial force resolution was set to an equivalent
Plummer gravitational softening of $15 \;h^{-1}$ comoving kpc. The SPH
smoothing length was set to the distance to the 40$^{th}$ nearest
neighbor of each SPH particle. In any case, we do not allow smoothing
scales to be smaller than the gravitational softening of the gas
particles. Using for three weeks 512 processors of the MareNostrum
supercomputer at BSC Barcelona (this time corresponds to 29 CPU years)
we have finished the simulation and created the {\em MareNostrum
  Universe}.

\begin{figure}
  \includegraphics[width=.5\textwidth]{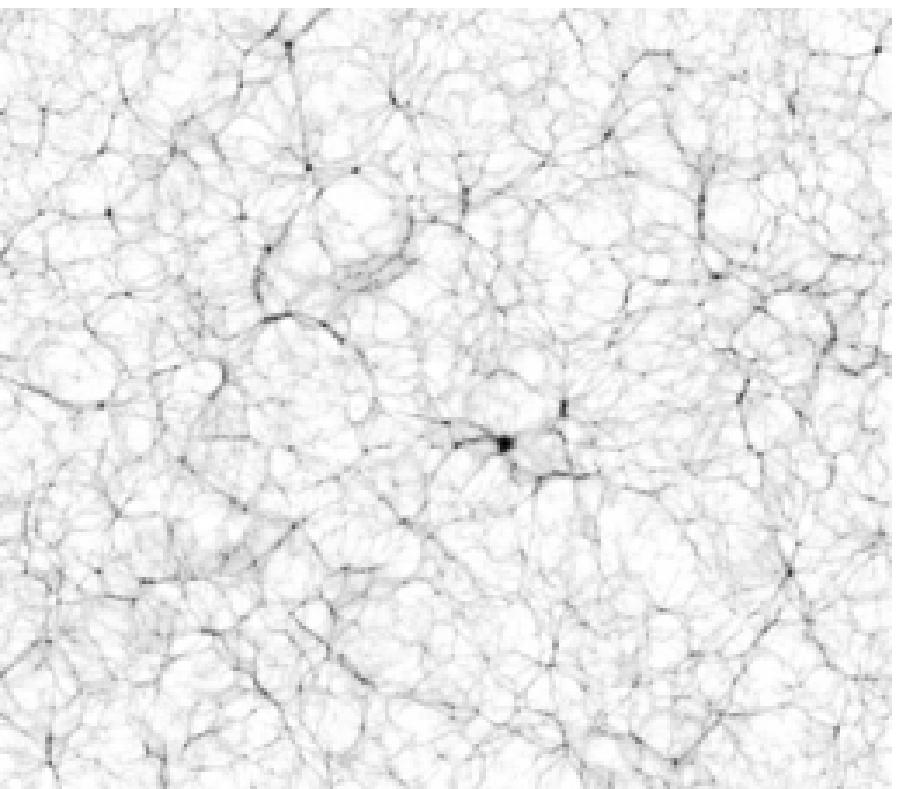}
  \includegraphics[width=.5\textwidth]{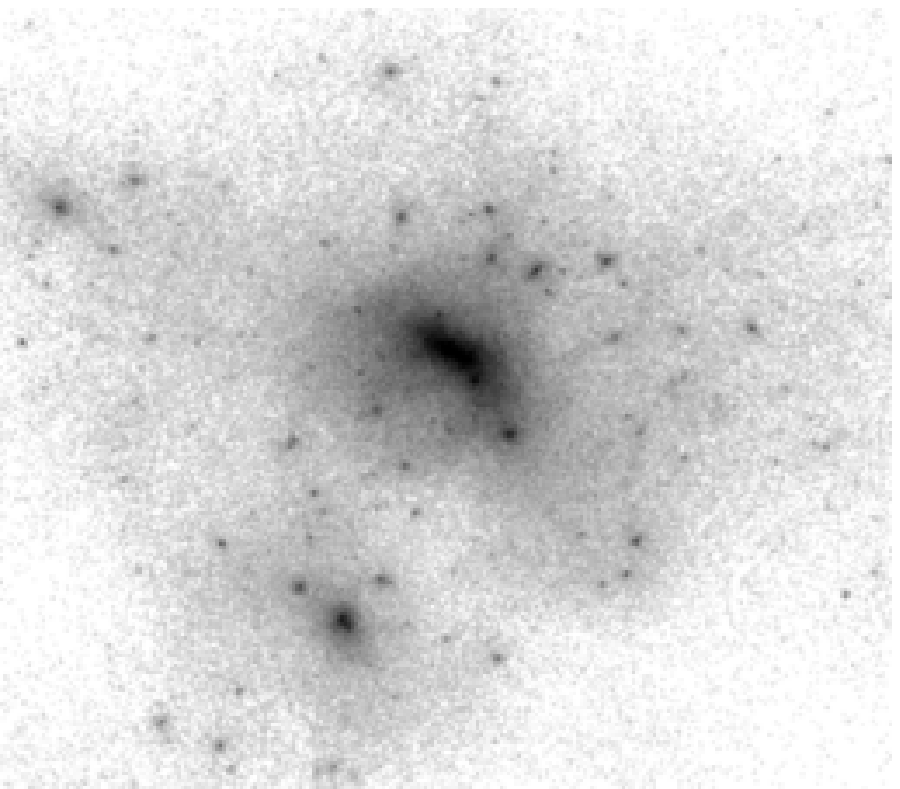}
  \caption{Left: The slice through the simulation contains the second most
    massive cluster found in the box at redshift $z=0$. Right: Zoom-in
    on this cluster. The size of the box is 20 $\hMpc$. Substructures
    can be clearly seen.}
\end{figure}

In Fig. 1 we show on the left a slice through the simulation showing
the density distribution of the dark matter component. One can clearly
see the large-scale filamentary structure. Approximately in the center
of the slice at the intersection of a few massive filaments one can
see the second most massive cluster in the simulation. It's total virial mass
(baryons and dark matter) is $2.2 \times 10^{15} \hMsun$ with a virial
radius of $2.6 \hMpc$. On the right hand side we show a zoom-in on
this cluster where the substructures can be seen.
\begin{figure}[h]
  \includegraphics[width=.5\textwidth]{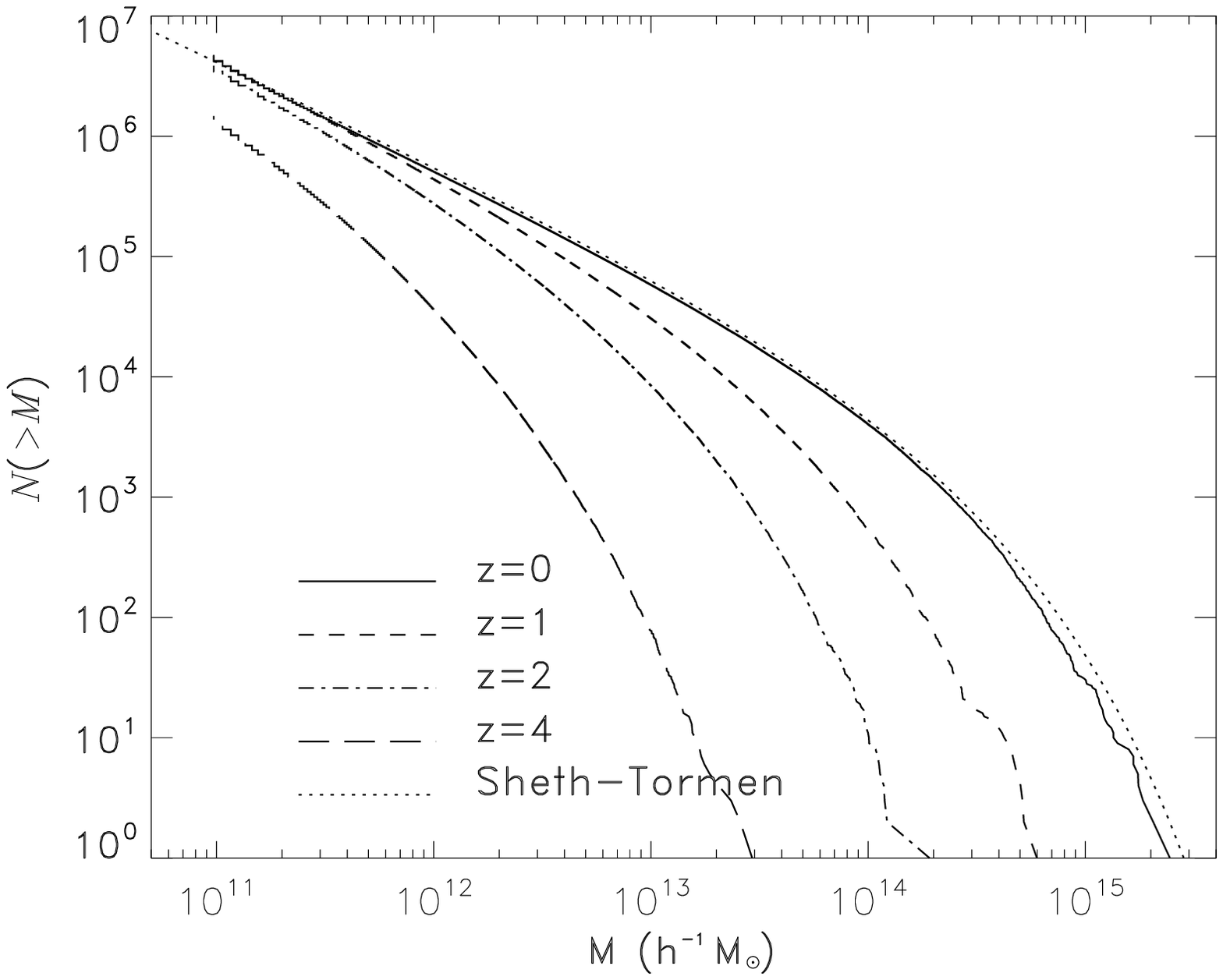}
\includegraphics[width=.5\textwidth]{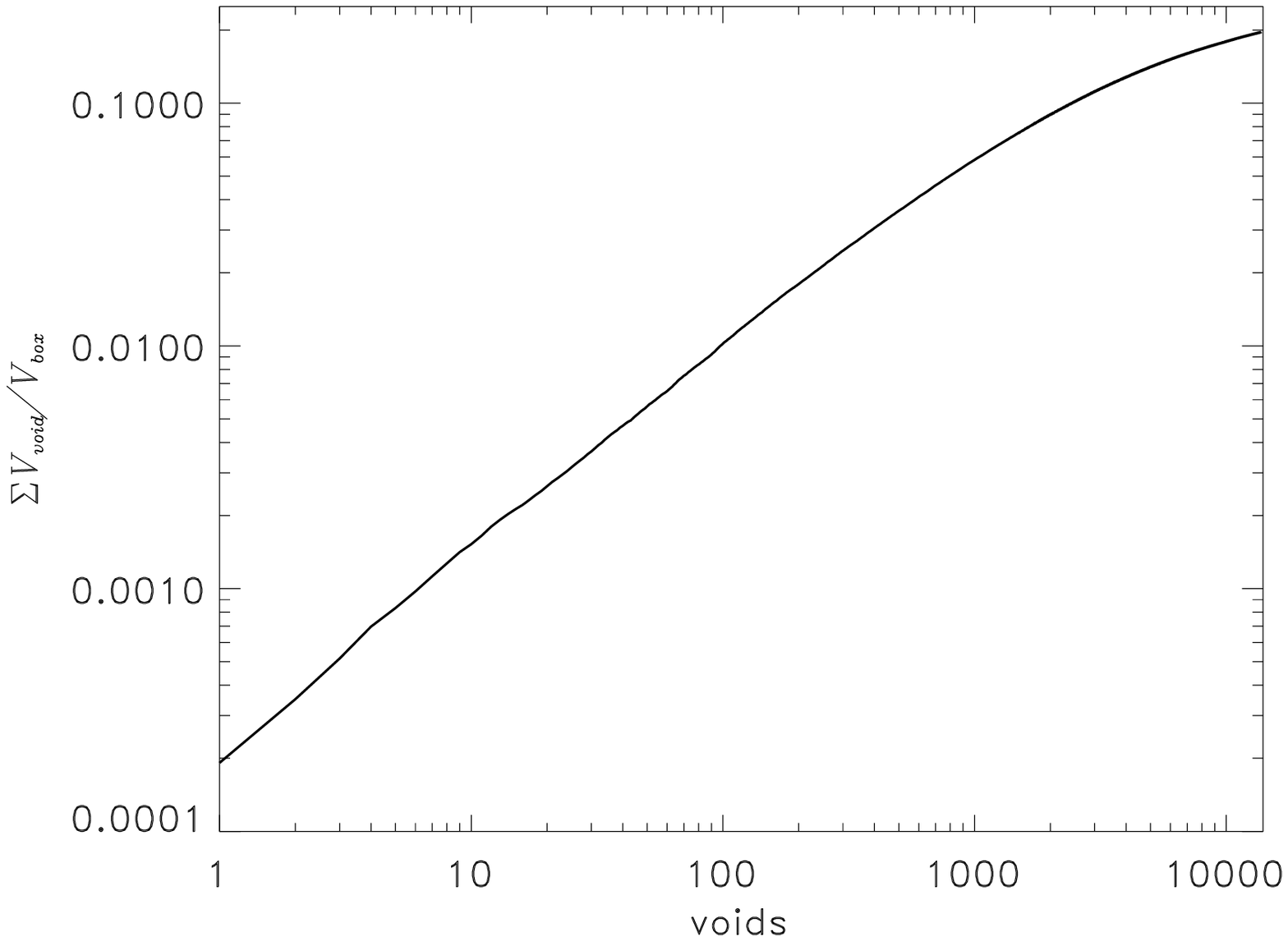}
  \caption{Left: Mass function of halos at redshift $z = 0, 1, 2, 4$ and
    Sheth-Tormen mass function (dotted line) at redshift $z=0$.
Right: Cumulative volume occupied by voids    }
\label{figtot}
\end{figure}

\subsection{Halos}

Gravitationally bound halos of different mass form has been formed
during the cosmological evolution. These halos consist of dark matter
and gas.  Due to the hydrodynamical interaction of the gas in general
the ration of gas to dark matter within the halos is different from
the mean ratio of 0.15 assumed in the simulation.  Moreover, the
spatial distribution of gas and dark matter differ inside the halos.
Even if one can see the halos my naked eye as white spots in Fig. 1 it
is a challenge to identify numerically all those halos within the
distribution of two billion particles and to determine their
properties.

For this purpose we have developed a parallel version of the
hierarchical friends-of-friends (FOF) algorithm described  in
\cite{klypin}.
 The FOF algorithm bases on the minimal spanning tree
(MST) of the particle distribution.  The minimal spanning tree of any
point distribution is a unique well defined quantity which describes
the clustering properties of the point process completely \cite{bahv}. 
The minimal spanning tree of $n$ points
contains $n-1$ connections. Based on the minimum spanning tree we sort
the particles in such a way that we get a cluster-ordered sequence
$P=\{p_1,p_2,...,p_n\}$. Any particle cluster is a segment of the
sequence $P$, i.e. it consist of points $p_ i,p_{i+1},...,p_j$ for
some indexes $i$ and $j$. Neighboring clusters, i.e. clusters which
merged immediately after increasing $r$, are neighboring segments on
$P$. Let us denote the length at which clusters $p_i, p_{i+1},...,
p_j$ and $p_{j+1},p_{j+2},...,p_k$ merge, by $r_{j+1/2}$. The
sequences $P$ and $R$ are sufficient for deriving the complete list of
clusters at any linking length $r$. In fact, the segment
$p_i,p_{i+1},..., p_j$ of the sequence $P$ is an $r$-cluster if and
only if $r_{i-1/2}>r$, $r_{j+1/2}>r$ and $r_{k+1/2}\leq r,\quad
k=i,i+1,...,j-1$. In other words, if all points would be located on a
line with distances $r_{j+1/2},j=1,2,...,n-1$ between neighboring
points, the line would break into the sequence of all $r$-clusters
after cutting of all segments larger than $r$. Obviously, the
sequences $P$ and $R$ (each of length $n_p \times 4$ byte) is the most
compact form to store the information about the whole hierarchy of
friends-of-friends clusters.

After topological ordering we cut the MST using different linking
lengths in order to extract catalogs of friends-of-friends particle
halos. Note, that cutting a given MST is also a very fast algorithm.
Typically we start with a linking length of 0.17 times the mean inter
particle distance which corresponds roughly to objects with the
virialization overdensity $\rho/\rho_{\rm mean}\simeq 330$ at $z=0$.
Decreasing the linking length by a factor of $2^n$ ($n = $1,2,...)we
get samples of objects with roughly $8^n$ times larger overdensities
which correspond to the inner part of the objects of the first sample.
With this hierarchical friends-of-friends algorithm we detect  all
substructures of halos. We are running the MST and FOF analysis
independently on the dark matter and gas particles to compare their
properties. At redshift $z=0$ we have detected 975500 objects with
more than 50 dark matter particles. All of these objects contain also
gas particles.  Running the MST and FOF analysis only over the gas
particles we have detected 630800 objects with more than 50 gas matter
particles which reflects the smoother distribution of the gas
particles.

In Fig. 2 we show the mass function of the FOF-halos at different
redshifts as total number of objects in the box.  Already at redshift
$z=2$ first few objects with masses larger than $10^{14} \hMsun$ have
been formed. 
 
\begin{figure}[t]
  \includegraphics[width=.5\textwidth]{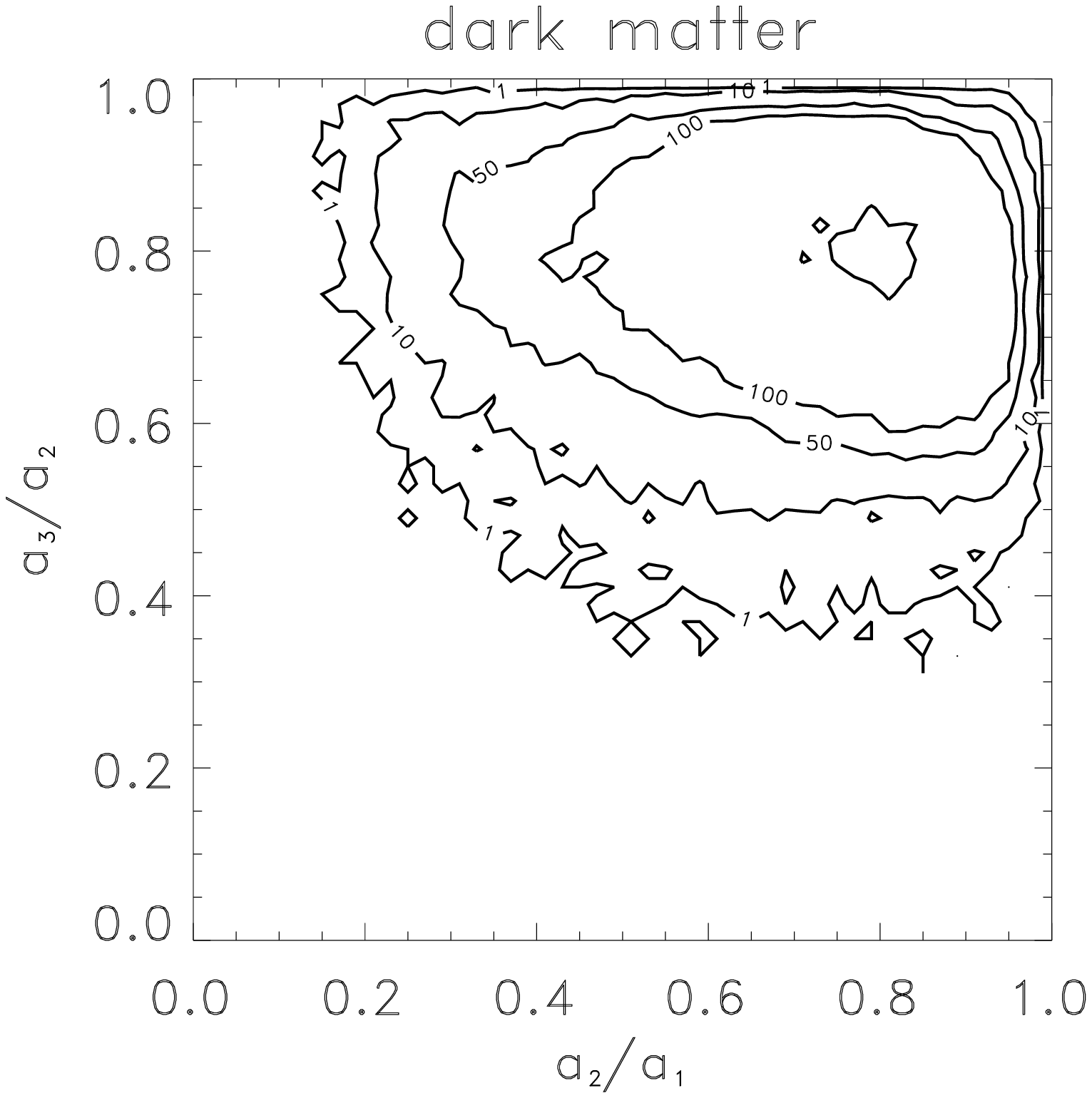}
  \includegraphics[width=.5\textwidth]{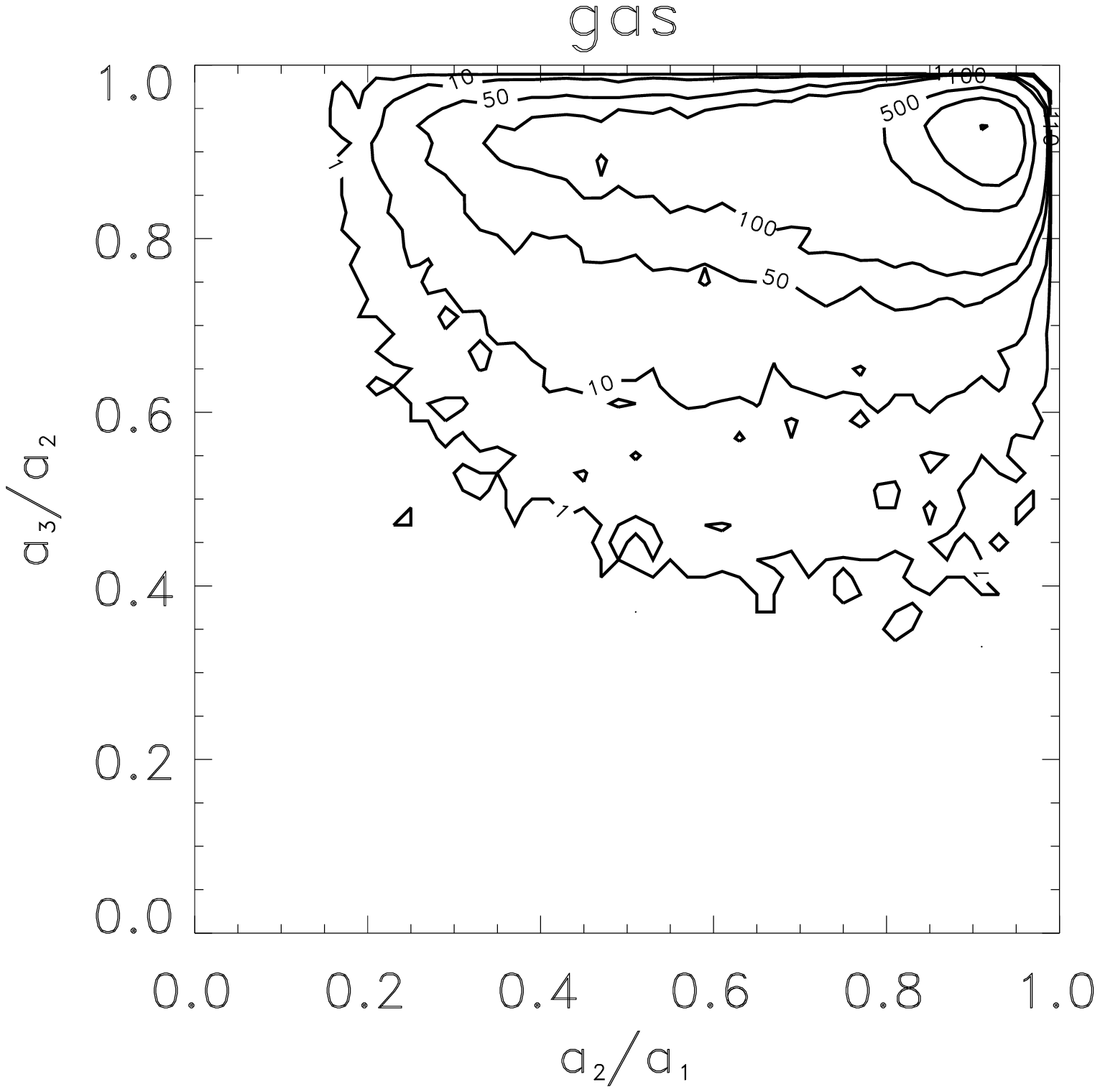}
  \caption{Left: Distribution of shapes of dark matter halos
    characterized by the ratio between main axes $a_2/a_1$ and
    $a_3/a_2$. The contouring is done by halo numbers per bin.
 Right: The same for the corresponding gas halos.}
\end{figure}
\subsection{Shape of Halos}

It is well known that dark matter halos have triaxial shapes and tend
do be prolate \cite{fal}. Their shape can be
characterized by the three eigenvectors of their inertia tensor. Thus
the shape of the objects is described by the three main axes $a_1 \geq
a_2 \geq a_3$ of a three-axial ellipsoid. In Fig. 3 we show the
distribution of shapes of halos with more than 500 particles. The
interval between 0 and 1 has been divided into 50 bins of length 0.02.
The contours show the number of halos per bin with the corresponding
ratios of axes.  Note, that the position of the maximum is mainly
determined by the large number of low mass halos.  In fact, due to the
power law of the mass function about 50 \% of the halos in this plot
have particle numbers between 500 and 1000 and more than 90 \% between
500 and 5000. The mean ratio of the axes of the dark matter and gas
halos depends on mass \cite{marenostrum}. Thus also the position
of the maximum in the shape distribution shown in Fig. 3 will depend
on the lower limit of mass assumed for the halos. However, the
qualitative behavior does not depend on mass. The dark matter halos
are triaxial and more prolate than oblate. The corresponding gas halos
are only slightly triaxial with much higher axis ratios, i.e. they are
more spherical.

\subsection{Voids}

In Fig. 1 (left panel) one can clearly see the filamentary structure
of the dark matter distribution with large empty regions between the
filaments. In the following we define voids as regions which do not
contain any halo more massive than $10^{12} \hMsun$. Voids in the
distribution of halos are found as described in \cite{voids}. In Fig.
\ref{figtot} (right panel) we show the cumulative volume occupied by
those voids.  The largest voids have radii of $18 \hMpc$ and thus
occupy only about about 0.02\% of the total volume. More than 14000
voids with radii larger than $5 \hMpc$ have been detected in the
distribution of 505539 halos with masses larger than $10^{12} \hMsun$.
They occupy a total volume of about 20 \% of the simulation box.

\begin{figure}[t]
  \includegraphics[width=.5\textwidth]{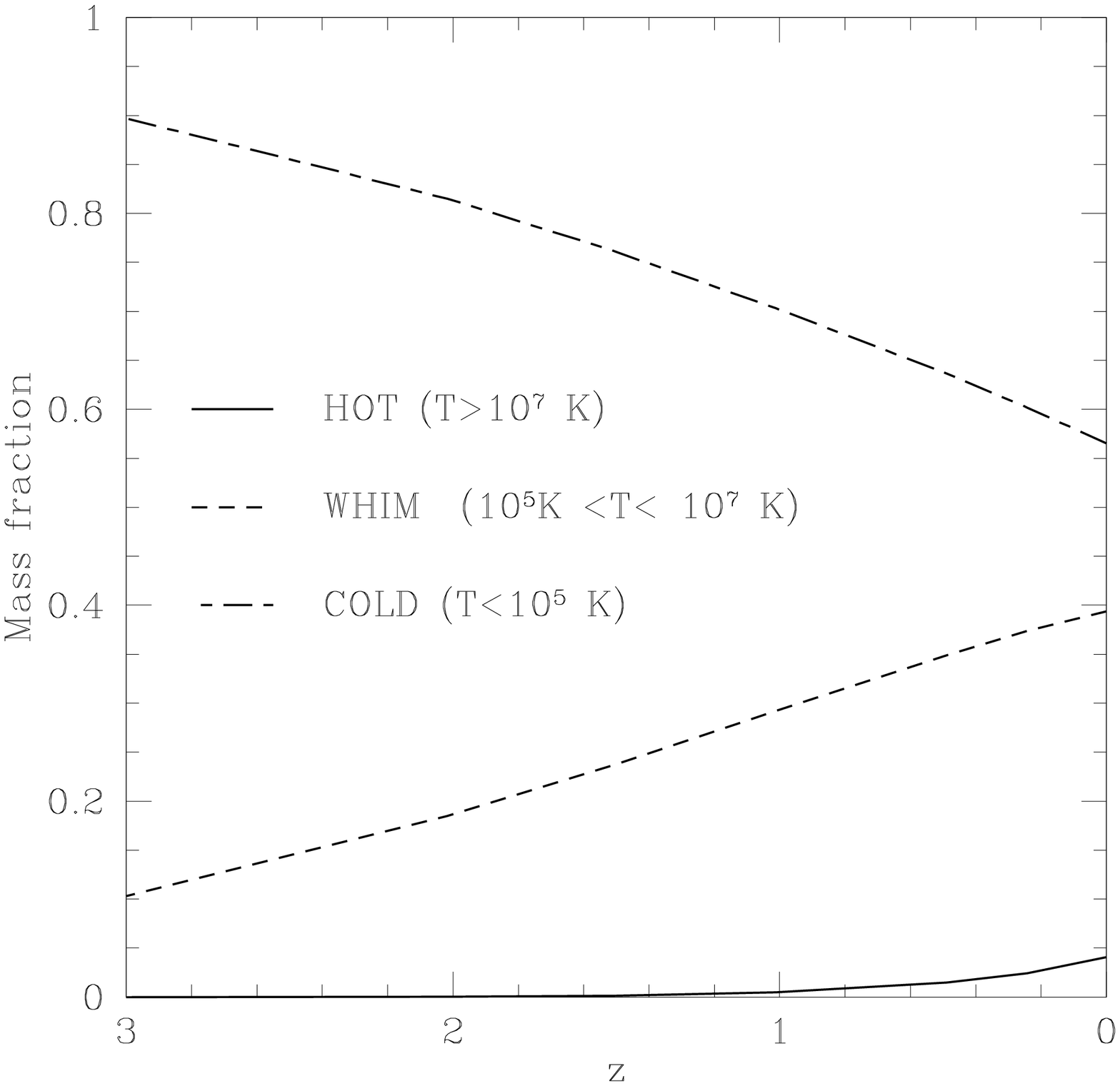}

  \includegraphics[width=.5\textwidth]{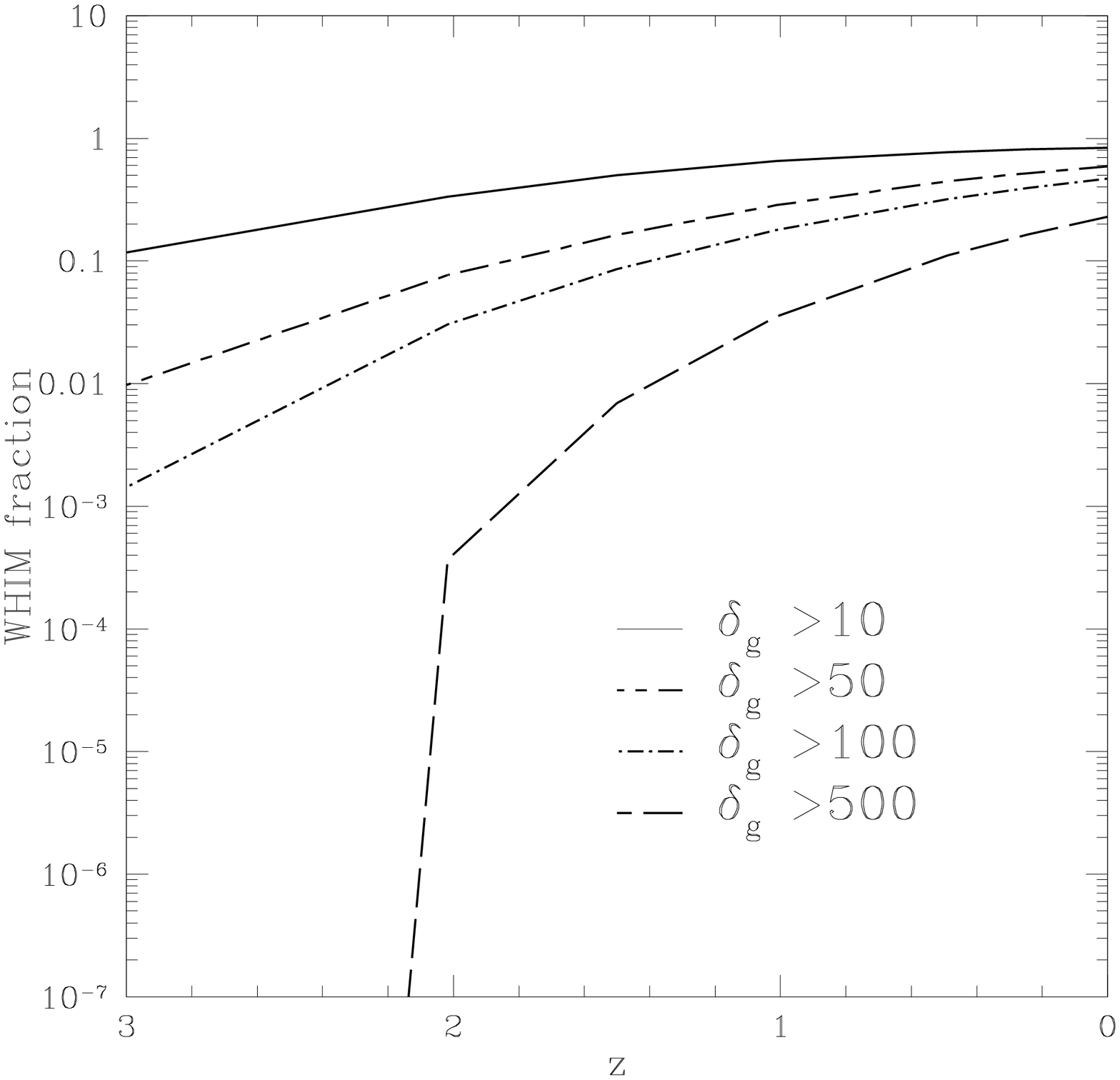}
  \caption{Left: Redshift evolution of the relative fractions of the 
 different baryon components selected according to the temperature of
 the gas particles. Right: Redshift evolution of the WARM-HOT baryon
 phase for different  overdensities.  
}\label{barion}
\end{figure}

\subsection{Baryon Distribution}

Due to the large number of gas particles that we have in this
simulation, we are able to trace accurately the time evolution of the
phase space ($T$--$\rho_{gas}$) distribution of baryons. We can then
study the redshift evolution of the amount of gas at different
temperatures inside the computational volume. In the left panel of
Fig. \ref{barion} we plot this evolution for 3 different baryon
components depending on the temperature: HOT ($T> 10^{7} K$), COLD ($T
< 10^{5} K$) and WARM-HOT ($10^5 K \leq T \leq 10^7 K $). As can be
seen in the figure, the amount of baryons in the WARM-HOT phase at
present correspond to roughly 40 \% of the total baryons in the
simulation. They would represent a substantial fraction of the
so-called missing baryons in the universe, as they would not be
detected in X-rays due to their relatively low temperatures.  In order
to see how much of these WARM-HOT gas is located outside the dark
matter halos, (\emph{i.e.}, populating the low density filamentary
structures or in voids), we show in the right panel of the same figure
the evolution with redshift of the WARM-HOT gas located inside
different local baryon overdensity thresholds. A baryon overdensity of
order 60 corresponds roughly with virial overdensities for an
isothermal sphere in the $\Lambda$CDM model ($\delta \sim 330$)
\cite{dave}. Therefore, we can see in the figure that the fraction of
baryons above certain overdensity increase rapidly at early times and
becomes stable once the universe is dominated by exponential
expansion ($z<0.5$). This is in good agreement with the assumption
that the baryons follow a log-normal density probability distribution
\cite{atrio}. We can also see from the figure that the amount of
WARM-HOT baryons that live in overdensities lower than those inside
virialized objects are of the order of 40\%.

\begin{figure}[t]
  \includegraphics[width=0.8\textwidth]{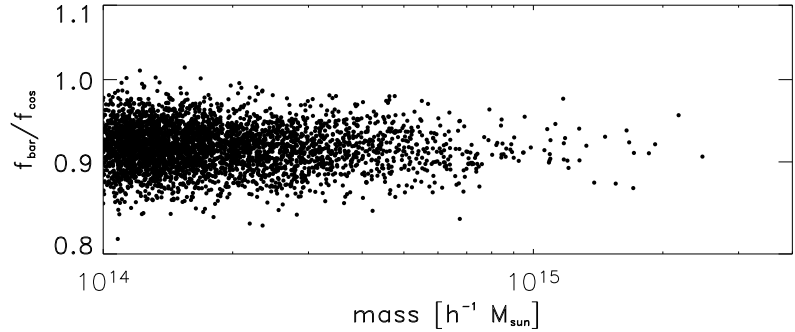}
  \caption{Baryon fraction in clusters
}\label{bfrac}
\end{figure}

One of the aims of performing this simulation was to obtain a large
database of galaxy clusters from which to study different
observational properties. In Fig \ref{bfrac} we plot the baryon
fraction for 4000 clusters in the simulation which have virial masses
larger than $10^{14} h^{-1} M_\odot $.  The baryon fraction is
normalized to the cosmic mean ($\Omega_B/\Omega_M=0.15$). As can be
seen, the total content of baryons inside clusters is always smaller
than the cosmic value for all clusters regardless of their mass.

\begin{theacknowledgments}
We would like to thank the Barcelona Supercomputer Center for allowing
us to run the simulation described above during the testing period of
MareNostrum. The analysis of this simulation has been done on NIC
J\"ulich. We also thank Acciones Integradas Hispano-Alemanas 
and DFG for supporting our collaboration.
\end{theacknowledgments}

\bibliographystyle{aipproc}

\begin{thebibliography}{10}
\bibitem{atrio} F. Atrio, J. M\"ucket, \emph{ApJ}, 643, 1 (2006)
\bibitem{bahv} S.~P. Bhavsar, R.~J. Splinter, \emph{MNRAS}\textbf{282},
  1461 (1996)
\bibitem{dave} Dav{\'e}, R., et al.\ 
 \emph{ApJ}, 552, 473 (2001)
\bibitem{fal} A. Faltenbacher, S. Gottl\"ober, M. Kerscher, V. M\"uller,
  V., \emph{A\&A}\textbf{387}, 778  (2002)
\bibitem{voids} S. Gottl\"ober, E. {\L}okas, A. Klypin, Y. Hoffman,
\emph{MNRAS}\textbf{344},  715 (2003)
\bibitem{marenostrum} S. Gottl\"ober, G.  Yepes, C.  Wagner, R. Sevilla,
  {astro-ph/0608289} (2006)
\bibitem{klypin} A. Klypin, S. Gottl\"ober, A.~V. Kravtsov, A.~M. Khokhlov,
  \emph{ApJS}\textbf{516}, 530 (1999) 
\bibitem{wmap} {D.~N. Spergel}, R. {Bean}, O. {Dore'}, M.~R {Nolta},
  C.~L. {Bennett}, G. {Hinshaw} et al.  {astro-ph/0603449} (2006)
\bibitem{gadget} V. Springel, \emph{MNRAS} 364, 1105 (2005)
\bibitem{millenium} V. Springel,  et al. \emph{Nature}, 435, 629 (2005) 

\end{thebibliography}

\end{document}